\documentclass[12pt]{article}
\usepackage{epsfig}
\usepackage{float}

\begin{document}
\rightline{NKU-2015-SF2}
\bigskip

\newcommand{\be}{\begin{equation}}
\newcommand{\ee}{\end{equation}}
\newcommand{\noi}{\noindent}
\newcommand{\ra}{\rightarrow}
\newcommand{\bib}{\bibitem}
\newcommand{\refb}[1]{(\ref{#1})}

\newcommand{\bff}{\begin{figure}}
\newcommand{\eff}{\end{figure}}

\begin{center}
{\Large\bf Quasinormal modes of  dilaton-de Sitter black holes: scalar perturbations}

\end{center}
\hspace{0.4cm}
\begin{center}
Sharmanthie Fernando \footnote{fernando@nku.edu}\\
{\small\it Department of Physics \& Geology}\\
{\small\it Northern Kentucky University}\\
{\small\it Highland Heights}\\
{\small\it Kentucky 41099}\\
{\small\it U.S.A.}\\

\end{center}

\begin{center}
{\bf Abstract}
\end{center}

Dilaton black hole solutions in low energy string theory  (well known as  GMGHS black holes) have  analogue black holes with a cosmological constant derived by Gao and Zhang. Here, we study  quasi normal modes of this dilaton-de Sitter black hole under neutral scalar field perturbations. We have employed the sixth order WKB analysis to compute the quasi normal mode frequencies.  A detailed study is done for the quasi normal mode frequencies by varying the parameters in the theory such as the mass, cosmological constant, dilaton charge and the spherical harmonic index. For the massive scalar field we observed that the usual quasi resonance modes that exists for asymptotically flat black holes do not exist for this particular black hole. We  have approximated the scalar field potential of the near-extreme dilaton de Sitter black hole with the P$\ddot{o}$schl-Teller potential and have presented exact quasi normal frequencies.

\hspace{0.7cm}


\section{ Introduction}

Recent observations of astronomical data indicates that the universe is expanding with an accelerated rate \cite{perl}\cite{riess}\cite{sper}\cite{teg}\cite{sel}. There are many proposals to explain the mysterious dark energy driving this acceleration of the universe. One of the simplest candidates for dark energy is the existence of the cosmological constant $\Lambda$ which was introduced first by Einstein to General Relativity in order to obtain a static universe.  The space-time with  a positive cosmological constant is known as de Sitter space-time. Hence, our universe might be described by a de-Sitter geometry.

In high energy physics, de Sitter spaces have taken an important role due to the dS/CFT correspondence; there is a holographic duality relating quantum gravity on the de Sitter space to conformal field theory on a sphere of one dimension lower \cite{witten}.  Another reason for de Sitter space-times to attract great deal of attention is  the need to understand de Sitter space-time in the context of string theory. If the real universe is de Sitter, then, a fully satisfactory de Sitter solution to string theory has to be found \cite{das} \cite{dan}.  In the low energy limit of  string theory, the Einstein action gets modified by scalar fields such as the dilaton and the axion  field. Hence, studies of black hole solutions with a dilaton field takes a important place in general relativity.

Black hole solutions with the dilaton field  have properties that are different from black hole solutions without it.   Earliest charged black hole solutions to dilaton gravity were found by Gibbons and Maeda \cite{gibbons}. It was also independently found by Garfinkle et.al. \cite{garf}.  This black hole is a solution to low energy string theory and is well known as the GMHGS black hole. 

First, let us present the action for dilaton gravity as,
\be
S = \int d^4 x \sqrt{-g } \left[ R - 2 \partial_{\mu} \Phi \partial^{\mu} \Phi - V(\Phi) - e^{- 2 \Phi} F_{\mu \nu} F^{ \mu \nu} \right]
\ee
Here $R$ is the scalar curvature, $F_{\mu \nu}$ is the Maxwell's field strength and $\Phi$ is the dilation field. The potential for the dilation field is given by $V(\Phi)$.

In an interesting paper, Poletti and Wiltshire \cite{pol3} considered a single Liouville-type potential for $V(\Phi)$, which is of the form, $ A e^{ a \Phi}$. They proved that if one expect asymptotically (anti)-de Sitter, spherically symmetric black hole solutions, that the potential $V(\Phi)$ has to be the one for the pure cosmological term  $ 2 \Lambda$.   In an interesting paper,  Gao and Zhang \cite{gao}, introduced three Liouville type potentials   given below to obtain non trivial exact  static spherically symmetric black hole solutions to dilaton gravity with a cosmological constant.
\be \label{potential}
V(\Phi) = \frac{ 4 \Lambda}{3} + \frac{ \Lambda}{3} \left( e^{ 2 ( \Phi - \Phi_0)} + e^{ - 2 ( \Phi - \Phi_0)} \right)
\ee
As one can see, the dilation potential is the sum of a constant and two Liouville type terms.

The potential used to obtain these solutions with a   combination of two or more exponential terms are not rare in physics literature. For example, for $N=4$ supergravity model presented by Gates and Zwiebach \cite{zwie1} \cite{ zwie2}, the potential was given as,
\be
V(\Phi) = C_1 e^{-2 \Phi} + C_2 e^{ 2 \Phi} + C_3
\ee
with $ C_1 = -\frac{1}{8}, C_2 = -\frac{\epsilon^2}{8}$ and $ C_3 = -\frac{\epsilon}{2}$.

In another  example, in  a paper by Easther \cite{easther}, number of exact solutions for the effective potential for the evolution of the Robertson Walker universe with a scalar field were derived. All of the solutions for the potential were of the form,
\be
V(\Phi) = \Sigma^N_{j}  \Gamma_j e^{ - \gamma_j \Phi}
\ee
New class of asymptotically AdS magnetic solutions in $(n+1)$ dimensions with three Liouville-type potentials for the dilaton were derived by Dehghani and Bazrafshan \cite{dehg}. Sheykhi et.al. \cite{she} derived black hole solutions to Einstein-Born-Infeld-dilaton gravity with a potential $V(\Phi) = 2 \Lambda_1 e^{ 2 \beta_1 \Phi} + 2 \Lambda_2 e^{ 2 \beta_2 \Phi}$. These black holes were non-flat asymptotically. Chan et.al. \cite{chan} derived charged dilaton black holes with unusual asymptotic with a potential similar to the one used by Sheykhi et.al.\cite{she}.

The metric of the  dilaton-de Sitter black hole derived by Gao and Zhang\cite{gao} is given by,
\begin{equation} \label{metric}
ds^2 = - f(r) dt^2 + \frac{ dr^2}{ f(r)} + R(r)^2 ( d \theta^2 + sin^2 \theta d \phi^2)
\end{equation}
where,
\begin{equation}
f(r) = 1 - \frac{ 2 M} { r} - \frac{ \Lambda r}{ 3} ( r - 2 D) ; \hspace{1 cm}R(r)^2 = r ( r - 2 D)
\ee
Here, $\Lambda$ is the cosmological constant, $M$ is the mass of the black hole and, $D$ is the dilation charge. 
Notice that when $\Lambda =0$, the black hole solution in eq.(\ref{metric}) becomes the well known GMGHS black hole \cite{gibbons} \cite{garf}. When $ D=0$, the space-time becomes the Schwarzschild-de Sitter black hole.  Hence,  Gao Zhang black hole given above is the best extension of the GMGHS to include a cosmological constant.

When a black hole is perturbed  by a field,  it undergo  damped oscillations  at the intermediate stage with complex frequencies. These oscillations are called quasi-normal modes (QNM) and their frequencies only depend on the parameters of the black hole.  Studies of QNM has attracted  many researchers in physics for variety of reasons.  From experimental point of view, if QNM are detected in gravitational antennas such as LIGO, VIRGO and LISA in the future \cite{ferra}, that will help identifying the physical properties of the black holes in the universe. QNM of asymptotically anti-de Sitter black holes have attracted lot of attention due to the famous AdS/CFT correspondence \cite{alex}. Another factor that has made QNM famous was Hod's conjecture \cite{hod} relating asymptotic values of  QNM frequencies  to quantum properties of black holes. There have been many works focused on computing asymptotic values of QNM frequencies due to this conjecture. Along those lines, many works also have obtained the area spectrum of quantum black holes \cite{fernando16} \cite{kun}.  More details on QNM can be found in the review by Konoplya and Zhidenko \cite{kono1}.

There are many works which have focused on QNM of dilaton black holes without $\Lambda$. Ferrari et.al. \cite{ferrari} studied gravitational perturbations. Scalar perturbations of dilaton black holes were done by several authors including Fernando and Arnold \cite{fernandok}, Konoplya \cite{kono7}\cite{kono8} and Shu and Shen \cite {shen}. Dirac field QNM were studied by Shu and Shen in \cite{shen}. Bifurcations of QNM spectrum of rotating dilatonic black holes were studied by Kokkotas et.al \cite{konobi}.

The paper is organized as follows: In section 2, an introduction to dilaton-de Sitter black hole  is given.  Basic equations for the scalar field around the black hole is presented in section 3. In section 4, QNM of massless scalar field is presented. In section 5, massive scalar field perturbations are presented. In section 6, the P$\ddot{o}$schl-Teller approximation is given. In section 7 the conclusion and ideas for future direction are given.


\section { Some properties of   the dilaton-de Sitter  black hole }

In this section we  will  discuss important properties related to the dilaton-de Sitter black hole in eq.$\refb{metric}$. First, notice that  the scalar curvature of the metric is,
\be
 \mathcal{R} = \frac{ 2 \left( - 2 D^3 r^2 \Lambda - 8 D r^4 \Lambda + 2 r^5 \Lambda + D^2 (r - 2 M + 9 r^3 \Lambda) \right)}{r^3 ( r - 2 D)^2}
 \ee
From the observation of the scalar curvature $\mathcal{R}$, it is clear that there is a singularity at $ r = 2D$.   The dilation field $\Phi$, the dilation charge $D$, and the electric field $F_{01}$, for the above solution are given by,
\be
e^{2 \Phi} = e^{ 2 \Phi_0} \left( 1 - \frac{ 2 D}{r} \right)
\ee
\be
D = \frac{ Q^2 e^{ 2 \Phi_0} }{ 2 M}
\ee
\be
F_{01} = \frac{ Q e^{ 2 \Phi_0}}{r^2}
\ee
$\Phi_0$ is the dilation field at $r \ra \infty$. $Q$ is the electric charge of the black hole.

The roots of the function $f(r)$  will give the horizons.  For small $M$ there are two horizons: one is the event horizon $r_h$ and the other, the cosmological horizon, $r_c$. For large $M$ there are no horizons and the space-time becomes a naked singularity. For a special value of $M$, the horizons become degenerate. Since the black hole has a singularity at $r = 2D$, the space-time becomes a black hole only if $r_h > 2 D$.

The Hawking temperatures of the black hole at the event and the cosmological horizon are given by,
\be
T_{h,c} = \frac{ 1}{ 6 \pi} \left| \frac{ 3 M}{ r_{h,c}^2} + ( D - r_{h,c}) \Lambda \right|
\ee


\section{ Basic equations for scalar perturbations}

The basic equation for a neutral scalar field is given by the Klein-Gordon equation, which is,
\be \label{klein}
(\bigtriangledown^{\mu} \bigtriangledown_{\mu} - m^2) \Psi = 0
\ee
The above equation can be separated to an angular part and a  radial part by substituting for $\Psi$ as,
\be
\Psi = e^{ - i \omega t} Y_{l,m} ( \theta, \phi) \frac{\Omega(r)}{R(r)}
\ee
Here, $Y_{l.m}(\theta, \phi)$ are the spherical harmonics. Once separated, the radial part will look like,

\be \label{wave1}
\frac{ d^2 \Omega(r_*) }{ dr_*^2} + \left( \omega^2 - V(r_*)  \right) \Omega(r_*) = 0
\ee
Here, $V(r_*)$  represents the effective potential for the scalar field given by,
\be
V(r_*) = f \left(\frac{ l ( l + 1) } { R^2} +  \frac{ f' R'}{ R} + \frac{ f R''}{ R}  + m^2 \right)
\ee
and $\omega$ is the frequency of the wave corresponding to the scalar field.   $r_*$ is the tortoise coordinate given by,
\be \label{tor}
dr_* = \frac{ dr} { f(r)}
\ee
eq.$\refb{tor}$ can be integrated to give,
$$
r_* = \frac{ - 3}{\Lambda ( r_c - r_h) ( r_h + r_v) ( r_c + r_v) } ( r_c ( r_h + r_v ) ln( r - r_c) - 
$$
\be
r_h ( r_c + r_v) ln ( r - r_h) - ( r_c - r_h) r_v ln( r + r_v) )
\ee
Here, $r_v$ is the third root of the function $f(r)$ which is unphysical since it is negative.\\

\noindent
When $r \ra r_h$, $r_* \ra - \infty$ and when $r \ra r_c$, $r_* \ra + \infty$.
The effective potential $V(r)$ depends on the parameters $ M, D, \Lambda$ and  $l$.  All potentials vanish at $r_h$ and $r_c$ and they are positive in between the horizons. 


\section{ Quasinormal modes for massless scalar perturbation}

In this section, we will focus on the QNM's of the dilaton-de Sitter black hole due to a neutral massless scalar field. First we will focus on the effective potential for $m =0$ field in the next section.


\subsection{ Effective potentials for a massless scalar field}

The effective potential is plotted for $ l=0$ and $ l=1$ for dilaton black holes with $ \Lambda =0$ and $\Lambda \neq 0$ in Fig.$\refb{potl}$. The figure on left represents the potential for $\Lambda \neq 0$. For $ l >0$, the potential is zero at $ r = r_h$ and $ r = r_c$ and positive in between.  For $ l =0$, the potential  has a local minimum between the two horizons and the minimum is negative. On the other hand the dilaton black hole with $\Lambda =0$ (shown on the right hand side) is positive for all $l$. This is a major difference between  the two black holes when it comes to scalar perturbations.

\begin{figure} [H]
\begin{center}
\includegraphics{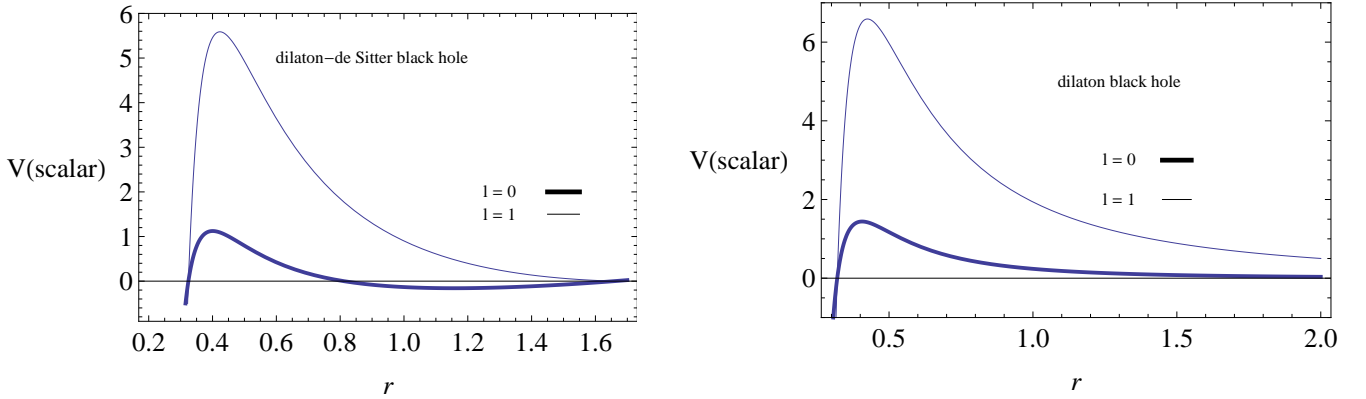}
\caption{The figure shows  $V(r)$ vs $r$ for  dilaton and dilaton-de Sitter black holes. Here $ M =0.16$, $\Lambda= 1, D = 0.1$}
\label{potl}
 \end{center}
\end{figure}


\subsection{ Computation of  QNM frequencies for $ l >0$ modes }

QNM are the solutions to the wave equation given in eq.$\refb{wave1}$ with pure ingoing wave at the black hole horizon. As for the boundary condition at  the cosmological horizon, the wave  will be purely out going. QNM frequencies will be complex and will be written as $\omega = \omega_R - i \omega_I$.  There are many methods developed to calculate QNM frequencies. The reader is referred to the excellent review by Konoplya  and Zhidenko \cite{kono1} for various methods of computation of QNM frequencies. Here, we will use the WKB approach developed by Iyer and Will \cite{will} and developed to sixth order by Konoplya \cite{kono4}. This method has been employed to compute QNM frequencies in  \cite{fernando6}\cite{fernando7}\cite{fernando8}. In this method. the QNM frequencies are related to the effective potential as,

\be
\omega^2 = - i \sqrt{ - 2 V_m''} \left( \Sigma^6_{i=2}  \Gamma_i  +   n + \frac{ 1}{2} \right) + V_m
\ee
Here, $V_m$ is the maximum value of the potential and $V_m''$ is the second derivative of the potential at the peak of the potential.  Expressions for $\Gamma_i$ can be found in \cite{kono4}. Here $n$ is the mode number. Since $n=0$ is the smallest frequency which leads to the largest time to damp the mode, the stability of the black hole will depend on the  fundamental frequency. Hence our major focus will be on  the fundamental frequency which has  $n =0$.

We have computed $\omega$ by varying the parameters in the theory, $M, D, \Lambda, n$ and $l$. All $\omega_I$ values were negative. We have used only the magnitude of  $\omega_I$ when plotting the graphs in the rest of the paper. First, let us study the dependency of $\omega$ with $D$ as shown in Fig.$\refb{omegad}$. Both $\omega_R$ and $\omega_I$ increases with $D$.  In particular, the damping is greater for greater $D$. Hence, the presence of the dilaton leads to more stable black holes. Therefore, dilaton-de Sitter black hole is more stable than the Schwarzschild de-Sitter black hole with the same mass.  In Fig.$\refb{omegacosmo}$, $\omega$ vs $\Lambda$ is plotted. Both $\omega_R$ and $\omega_I$ decreases when $\Lambda$ increase. Hence, smaller $\Lambda$ is preferred for stable black holes. In Fig.$\refb{omegasphe}$,  $\omega$ vs l (the spherical harmonic index) has been plotted for $n =0$. It is clear that there is a linear relation between $\omega_R$ and $l$. $\omega_I$ decreases with increasing $l$ and then reach a constant value for larger $l$. It was observed that the temperature depend on  $\omega_I$  linearly for asymptotically anti de Sitter black holes $\cite{horo11}$. We plotted $\omega_I$ vs  temperature for the dilaton-de Sitter black hole in Fig.$\refb{omegatemp}$ and saw that in fact the two quantities relate to each other  linearly.

\begin{figure} [H]
\begin{center}
\includegraphics{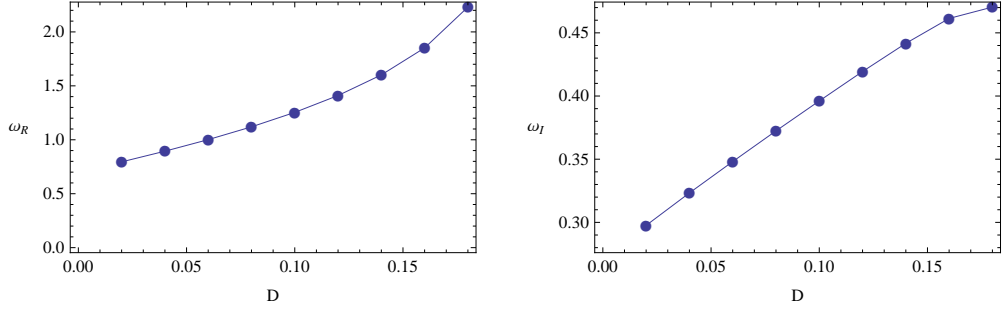}
\caption{The figure shows the $\omega$ vs $D$.  Here, $M = 0.2, l = 1$ and $ \Lambda = 2$.}
\label{omegad}
 \end{center}
\end{figure}

\begin{figure} [H]
\begin{center}
\includegraphics{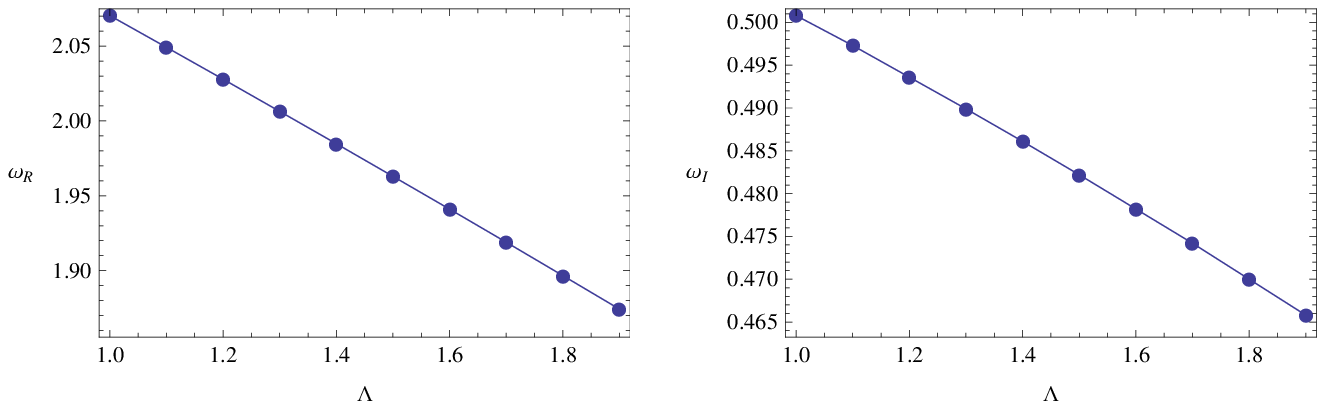}
\caption{The figure shows the $\omega$ vs $\Lambda$.  Here, $M = 0.2, l = 1$ and $ D = 0.16$.}
\label{omegacosmo}
 \end{center}
\end{figure}

\begin{figure} [H]
\begin{center}
\includegraphics{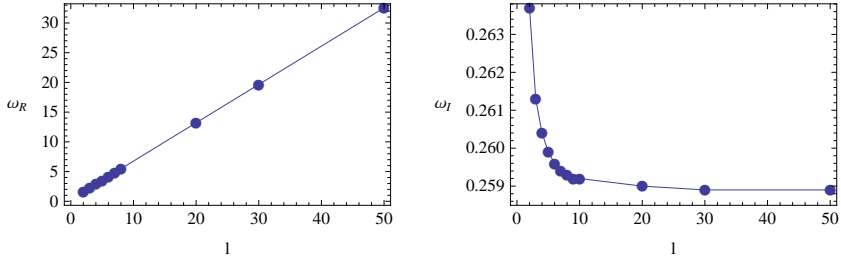}
\caption{The figure shows the $\omega$ vs $l$.  Here, $M = 0.25, \Lambda = 2$ and $ D = 0.16$.}
\label{omegasphe}
 \end{center}
\end{figure}

\begin{figure} [H]
\begin{center}
\includegraphics{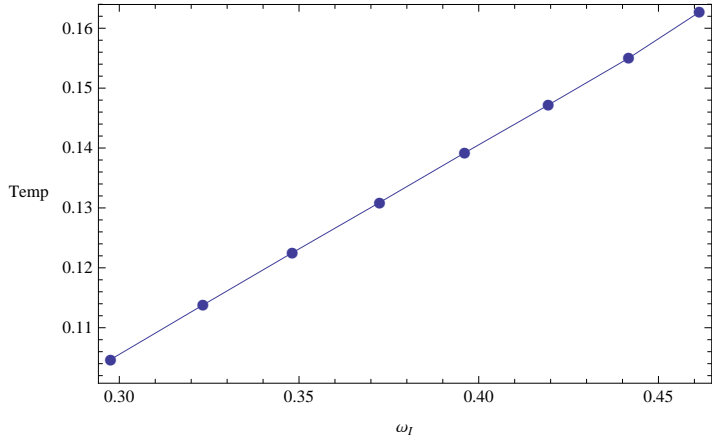}
\caption{The figure shows the $Temp$ vs $\omega_I$.  Here, $l=1, \Lambda = 2$ and $M = 0.2$.}
\label{omegatemp}
 \end{center}
\end{figure}


\subsection{ Scalar perturbation of $l =0$ modes }

As shown in Fig.$\refb{potl}$, the effective potential for $ l=0$ mode significantly differs for the $ l>0$ modes. It is negative in some region between $r_h$ and $r_c$. Similar behavior for $ l=0$ mode potential has been shown  for the Reissner-Nordstrom-de Sitter black hole \cite{brady} and for the regular-de Sitter black hole \cite{fernando9}. Due to the fact that the potential has an extra local minimum, WKB approximation is not appropriate to compute frequencies. If one needs to compute QNM frequencies, a time-domain integration method introduced by Gundlach et.al \cite{gund} can be employed. In this paper, instead of computing the QNM frequencies, we will analytically investigate if $l=0$ mode decays or not.

First let us introduce null coordinates $u$ and $v$ as,

\be
u = t - r_*; \hspace{1 cm} v = t + r_*
\ee
According to the null coordinates, the future black hole horizon $r_h$ is located at $ u = \infty$ and the future cosmological horizon $r_c$ is located at $ v = \infty$. The equation for the scalar field in the new coordinates is,
\be \label{wave3}
\frac{ \partial^2 \eta(u,v)}{\partial u  \partial v} = - \frac{1}{4} V_{0}(r) \eta(u,v)
\ee
Here,
\be
\Psi(u,v, \theta, \phi) =  Y_{l,m}(\theta, \phi) \frac{ \eta(u,v)}{R(r)}
\ee
and,
\be
V_{0} =\frac{ f (fR')'}{R}
\ee
The solution to eq.$\refb{wave3}$ can be expanded in a series of two arbitrary  functions, $G(u)$ and $K(v)$ as \cite{brady} \cite{gund},

\be \label{expan}
\eta(u,v)= a_0 \left( G(u) + K(v) \right) +  \sum_{z=0}^{\infty} B_z(r) [ G^{(-z-1)} (u) + (-1)^{z+1} K(v)^{(-z-1)} ]
\ee
Negative super indices on $G(u)$ and $K(v)$ refer to integration with respect to $u$ and $v$ respectively: for example, $K^{-1}(v) = \int K(v) dv$. The coefficient $a_0$ can be set equal to $1$ without loss of generality. By substituting  eq.$\refb{expan}$  to eq.$\refb{wave3}$, one can  obtain  recurrence relation for $B_z(r)$ as,
\be \label{recur}
B_{z+1}' =     - \frac{ ( f R')'}{ 2 R} B_z + \frac{ f'}{2}  B_z'   +  \frac{ f}{2} B_z''
\ee
Here,
\be \label{bnotee}
B_0(r) = - \int \frac{(f R')'}{2 R} dr
\ee
after integration, it is given by,
\be \label{bnote}
B_0(r) = \frac{ 3 M}{ 8 r^2} - \frac{ 1}{ r} - \frac{ ( D - M)}{ 8  D ( - 2 D + r)} + \frac{ r \Lambda}{3} + \frac{ ( 6 D - 3 M - 4 D^3 \Lambda)}{ 48 D^2}   ln \left( \frac{ r}{ - 2 D + r} \right)
\ee
One can compute other $B_z$ values with the use of the recurrence relations.  

Now, let us consider an initial burst of radiation at $ \nu =0$ which is confined between $ u =0$ and $ u = u_1$. This burst is given by $\eta(u, 0) = G(u)$ and $ \eta(0, \nu) =0$. Fig.$\refb{burst}$ demonstrate the location of null coordinates and the initial burst as well as the horizons.

\begin{figure} [H]
\begin{center}
\includegraphics{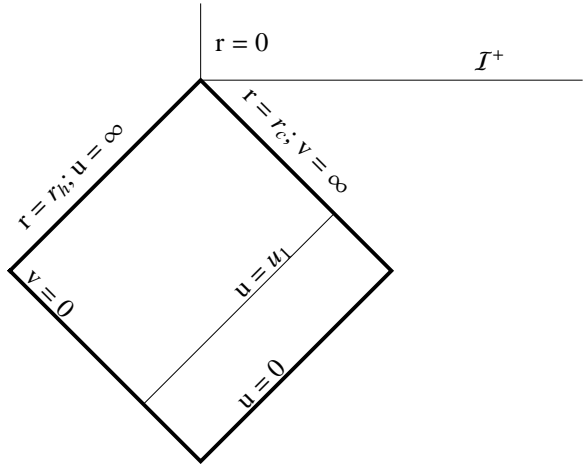}
\caption{The figure shows the location of null coordinates for the dilaton-de Sitter black hole. The  initial burst is confined between $u=0$ and $u=u_1$ and on $v=0$. The initial burst is such that, $\eta(u,0) = G(u)$ where $G(u) \neq 0$, for $0<u<u_1$ only.}
\label{burst}
 \end{center}
 \end{figure}
 
Following the work by Gundlach et.al \cite{gund} and Brady et.al. \cite{brady}, the evolution of the scalar field of the initial burst can be studied in   two parts:
\noindent
(i) the evolution of the burst $\eta(u, 0) = G(u)$ which is non-zero between $0<u<u_1$ and $\eta(0,v) = 0$.
(ii) subsequent evolution in the region for $u \geq u_1$.


\subsubsection{ The evolution of the burst $\eta(u, 0) = G(u)$ which is non-zero between $0<u<u_1$ and $\eta(0,v) = 0$ }

Since the initial burst is independent of $v$ and $G(u_1) =0$, the field at $u = u_1$ is given by,
\be \label{A4}
\eta(u_1, v) = \sum_{z=0}^{\infty} B_z(r) G(u)^{(-z-1)} (u_1)
\ee
For sufficiently small $\Lambda$, the cosmological horizon $r_c$ is large. When one observe $B_z$ values for small $\Lambda$ and large $r_c$, it is clear that except $B_0$, all other $B_z$ vales approximate zero. This is due to the fact that $B_z$ for $ z >0$ has $\Lambda^2$ terms or higher order terms or $r^{-1}$ or higher order terms.  Hence, at the cosmological horizon $r_c$ ($ v = \infty$),

\be \label{A5}
\eta(u_1, \infty) \approx  B_0(r_c) G^{-1}(u)  = B_0(r_c) \int^{u_1}_{0}  G(u) du + O(\Lambda)
\ee


\subsubsection{ The evolution of the field in $ u  \geq u_1$ region}

Now, one can examine the evolution of the field in the region $ u  \geq u_1$ by taking the field $\eta(u_1,v)$ in eq.$\refb{A4}$ as the initial data for small $\Lambda$. The field equation given in eq.$\refb{wave3}$ can be solved near the cosmological horizon as \cite{brady} \cite{gund},
\be \label{A52}
\eta(u,v) \approx [ G(u) + K(v) ]  + B_0(r) [ G^{-1}(u) - K^{-1}(v) ]
\ee
Defining $G^{-1}(u) = g(u)$ and $K^{-1}(v) = -k(v)$ where $g(u)$ and $k(v)$ are arbitrary functions, one can rewrite eq.$\refb{A52}$ as,
\be \label{A6}
\eta(u,v) \approx \ B_0(r) [ g(u) + k(v) ] + [ \frac{ \partial g(u)}{ \partial u} - \frac{ \partial k(v)}{ \partial v} ]
\ee
From eq.$\refb{tor}$,  closer to the cosmological horizon $r_c$, $r_* \approx -\frac{1}{f'(r_c)} Log( r_c - r)$. Hence,   given  the fact that $ r_*= \frac{ v- u}{2} $, one can write $r$ closer to $r_c$  as,
\be \label{newr}
r \approx r_c - e^{  -\kappa_c v} e^{ \kappa_c u}
\ee
The parameter  $\kappa_c$ above is the  surface gravity at $r_c$. Hence closer to the cosmological horizon ( i.e. $v \ra \infty$) the right hand side of eq.$\refb{A4}$ can be expanded as a polynomial of $e^{- \kappa_c v}$. By comparing eq.$\refb{A4}$ and eq.$\refb{A6}$ one can conclude that the function  $k(v)$ is also a polynomial of $e^{- \kappa_c v}$ given as,
\be \label{A7}
k(v) = \sum_{n=0}^{\infty} k_n e^{- n \kappa_c v}
\ee
When $ v \ra \infty$,  $k_0$ can be computed from eq.$\refb{A5}$ and eq.$\refb{A6}$,  as,
\be
k_0 = \frac{1}{ B_0(r_c)} \eta(u_1, \infty)
\ee
which is non-zero.
Even though the nature of the function $g(u)$ depends on the potential everywhere,  since $g(u)$ also evolves from eq$\refb{A4}$, it seems reasonable to predict that it is of the form (when $ u \ra \infty$)
\be
g(u) = \sum_{n=0}^{\infty} g_n e^{- n \kappa_c u}
\ee
Hence, at late times, the  field $\eta \ra k_0 + g_0$ and the scalar field $\Psi =  \frac{ \eta}{R(r)} = \frac{ k_0 + g_0}{R(r_{c,b})}$ approach a constant value(when $u$ and $v$ both approach $\infty$).  For  $l=0$, this is the only static solution which is regular at both horizons $r_b$ and $r_c$. Hence, the field decay into a constant value at late times for $ l=0$. In contrast modes for   $l>0$  decay to zero. 

Given the observation that $l=0$ mode reach a constant value according to the analysis presented here, the question arises whether the black hole is stable or not for $l=0$ modes. As presented in Fig.$\refb{potl}$, the effective potential for $l=0$ modes has a negative region in contrast to $ l >0$ modes. Even though this may suggest that the black hole may be unstable for $ l=0$ mode,  this is in fact not true for some potentials. For example, Bronnokov et.al \cite{kono15} analyzed wormholes with a phantom scalar field for $l=0$ mode which had a potential with a negative region, and, demonstrated that the field in fact decay without causing instability.  Therefore if one needs to conclude the status of the stability of the black hole considered in this paper for $l=0$ mode, then a more thorough analysis of the quasi normal mode behavior has to be done. In other words one has to study the  time-domain profile for the scalar mode behavior using a finite difference technique done in \cite{gund}. Here, in this paper we have omitted such an exhaustive analysis of the $l=0$ mode behavior and just predict the possibility of the mode to be stable.


\section{ Massive scalar perturbations and QNM }

In asymptotically flat space-times, such as the Schwarzschild black hole,  massive scalar field has been shown to decay slower than the massless scalar field \cite{kono5}. When the mass of the field is increased to  sufficiently large values, $\omega_I$ decreases to zero. Such modes with $\omega_I=0$ are called quasi-resonance-modes(QRM). Studies of QRM were first done by Ohashi and Sakagami for the Reissner-Nordstrom black hole in \cite{ohashi}.

The effective potential $V$ for the dilaton-de Sitter and the dilaton black holes are   plotted  in Fig.$\refb{potmass}$ by varying the mass $m$. Here, $l=1$. For the dilaton-de Sitter black hole, the potential height increases when  the mass is increased. Also, the potential is zero at $ r = r_h$ and $r= r_c$. For the GMGHS black hole (dilaton black hole with $\Lambda =0$), the  maximum of the potential increases when the mass $m$ is increased and for a critical value of the mass $m$, the peak of the potential ceases to exist. Also, the potential reaches a constant value when $r$ gets large.

\begin{figure} [H]
\begin{center}
\includegraphics{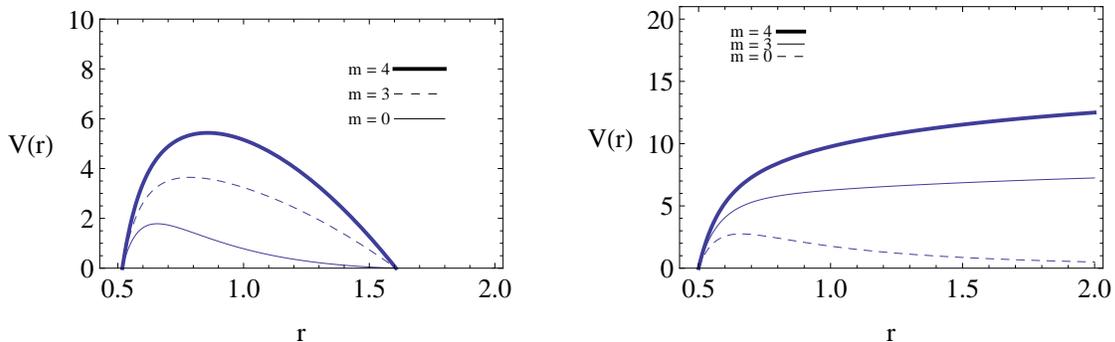}
\caption{The figure shows the $V(r)$ vs $r$ for the dilaton-de Sitter(left) and the dilaton (right) black holes.  Here, $M = 0.25, \Lambda = 1,l=1$ and $ D = 0.16$.}
\label{potmass}
 \end{center}
\end{figure}

First we want to discuss QNM for the massive fields when $ l >0$. We have presented QNM values for $ l=1$   in Table 1. Here, $\omega_R$ increases with $m$ and $\omega_I$ decreases with $m$. Hence, the massless field decay faster compared to the massive scalar field.  From Table 1,  one can observe that when $m$ increases for large values, $\omega_I$ decreases but does not approach zero. Notice that we have increased the mass $m$ upto the mass of the black hole. Hence, there are no QRM in  the dilaton-de Sitter black holes. This is similar to what was observed for the QNM of the massive scalar field of the Schwarzschild-de Sitter black hole  by  Chang et.al \cite{shen4}. The  reason why the de-Sitter black holes does not have QRM values is the fact that the boundary condition at $r= r_c$ is not fulfilled for the asymptotically flat black holes which is clear from Fig.$\refb{potmass}$.

\begin{center}
\begin{tabular}{|l|l|l|r|} \hline \hline 
 m & $\omega_R$  &  $\omega_I$  \\ \hline

0.25 &  0.472589 & 0.157059 \\ \hline
0.26 &   0.472987 & 0.156985\\ \hline
0.27 &   0.473401 & 0.156908 \\ \hline
0.28&   0.473830 & 0.156829 \\ \hline
0.29 &  0.474275 & 0.156747  \\ \hline
0.30 &   0.474735 & 0.156663 \\ \hline
0.31 &  0.475211  & 0.156576 \\ \hline
0.32 &   0.475702 & 0.156487 \\ \hline
0.33 &   0.476208 & 0.156396 \\ \hline
\end{tabular}
\end{center}

 Table 1: QNM frequencies corresponding to the  mass of the scalar field  $m$. Here $ M = 0.34, \Lambda =1, D = 0.14$ and  $  l = 1  $

For $ l=0$, the potential for the massive field is plotted in Fig.$\refb{potmass2}$. Again, the dilaton-de Sitter black hole potential shows similar behavior as the massless potential for $l=0$. However, for large mass $m$, the potential will be positive between the horizons similar to $l=0$ case. When the potential has a negative region, one can do a similar analysis as done for the massless field in section (4.3) which would lead to a field reaching a constant at the cosmological horizon.

\begin{figure} [H]
\begin{center}
\includegraphics{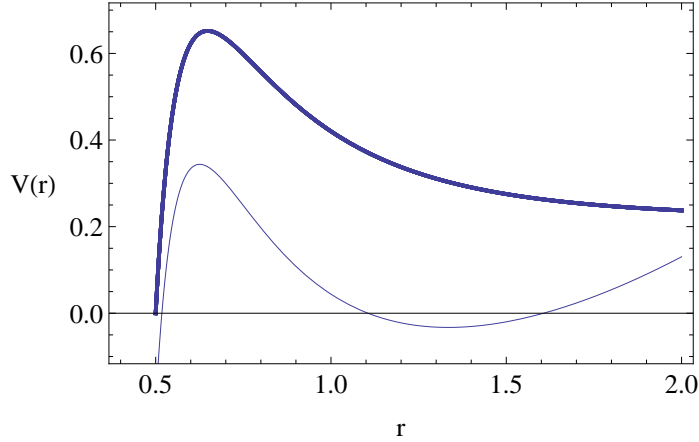}
\caption{The figure shows the $V(r)$ vs $r$ for the dilaton-de Sitter(thin curve) and the dilaton (thick curve) black holes.  Here, $M = 0.25, \Lambda = 1,l=0$ and $ D = 0.16$.}
\label{potmass2}
 \end{center}
\end{figure}


\section{ P$\ddot{o}$schl-Teller approximation for the near-extreme dilaton-de Sitter black hole}

First we would like to present some properties of the extreme dilaton-de Sitter black hole as follows:

\subsection{ Extreme dilaton-de Sitter black hole}

Here we consider extreme black holes when $ r_h = r_c$. In that case, two conditions has to be satisfied:
\be 
f(r) =0;  \hspace{ 1 cm}  f'(r) =0
\ee
When the dilaton-de Sitter black hole is extreme, $r_{extreme} = r_h = r_c = \rho$ is given by,
\be \label{rextreme}
\rho = \frac{  2 D + \sqrt{ 4 D^2 + \frac{9}{ \Lambda}}}{3}
\ee
One can observe that when $ D \ra 0$, $ \rho \ra \frac{3}{\sqrt{\Lambda}}$ which is the value for the extreme radius of the Schwarzschild-de Sitter black hole \cite{pod}. 

The function $f(r)$ for the extreme black hole is,
\be
f(r) = \frac{ -\Lambda ( r -\rho)^2 ( r + b)}{ 3 r}
\ee
where
\be
b = \frac{ 6 M_{critical}}{ \Lambda \rho^2}
\ee
Here, $M_{critical}$ is the mass of the black hole when the horizons are degenerate.
Then,
\be
f''(\rho) = \frac{ d^2f(r)}{dr^2} =  -\left(\frac{ 4M_{critical}}{\rho^3} + \frac{ 2 \Lambda}{ 3} \right)
\ee
Such extreme black holes where $r_h = r_c$ are called Nariai black holes. Nariai black holes in variety of context have been studied in \cite{fernando3}\cite{fernando4}\cite{fernando5}.

\subsection{ Near-extreme black hole}

In the near-extreme black hole, $r_h$ is very close to $r_c$.  Hence one can expand $f(r)$ in a Taylor series as \cite{paw},
\begin{equation}
f(r) \approx  \frac{ f''(\rho)}{2} ( r - r_h) ( r - r_c)
\end{equation}
Hence the tortoise coordinate defined in eq.$\refb{tor}$ can be integrated as,
\be \label{tor2}
r_* = \int \frac{dr}{ f(r)} =   \frac{2}{ f''(\rho) ( r_c - r_h)} log\left( \frac{ r_c - r}{ r - r_h} \right) 
\ee
When $r \ra r_h$, $r_* \ra -\infty$ and $ r \ra r_c$, $r_* \ra \infty$. 
\noindent
From the relation in eq.$\refb{tor2}$
\be \label{rvalue}
r = \frac{ r_h + r_c e^{ \eta r_*}}{ 1 + e^{ \eta r_*}}
\ee
Here,
\be \label{eta}
\eta = -\frac{f''(\rho)}{2}  ( r_c - r_h) 
\ee
With the  definition given in eq.$\refb{rvalue}$ for $r$, the function $f(r)$ for the near extreme dilaton-de Sitter black hole can be written as,
\be \label{newfr}
f(r)  =  \frac{ \gamma}{ 4( Cosh( \frac{ \eta r_*}{2}))^2} 
\ee
Here,
\be
\gamma =  -\frac{ f''(\rho)}{2} ( r_c - r_h)^2
\ee
Now the wave equation for the scalar field perturbation can be written as,
\be \label{wave}
\frac{ d^2 \Omega(r_*) }{ dr_*^2} + \left( \omega^2 - V_{eff}(r_*)  \right) \Omega(r_*) = 0
\ee
with,
\be
V_{eff} =  \frac{ V_0}{ Cosh^2( \frac{\eta r_*}{2})}
\ee
where
\be \label{v0}
V_0 = \frac{ \gamma l ( l+1)}{ 4 \rho^2}  + \frac{\gamma}{4} m^2
\ee
In deriving $V_0$,  we have assumed $ 2 D \ll r_h$. Hence $R(r) \approx r$, $R''(r) \approx 0$ and $R(\rho) \approx \rho$. Also since $f(r)' \approx 0$, only the first term in the potential dominates.

Now the wave equation for the scalar field perturbations simplifies to,
\be \label{wave2}
\frac{ d^2 \Omega(r_*) }{ dr_*^2} + \left( \omega^2 -   \frac{ V_0}{ Cosh^2( \frac{\eta r_*}{2})}  \right) \Omega(r_*) = 0
\ee
$\frac{ V_0}{ Cosh^2( \frac{\eta r_*}{2})}$ is the well known   P$\ddot{o}$schl-Teller potential. Ferrari and Mashhoon,  in a well known paper \cite{ferra3} demonstrated that  $\omega$ can be computed exactly as,
\be
\omega =  \sqrt{ V_0  - \frac{\eta^2}{16}} - i \frac{\eta}{2} ( n + \frac{1}{2})
\ee
The P$\ddot{o}$schl-Teller approximation was used to obtain exact expressions for QNM of scalar, electrodynamic and gravitational perturbations of a near extreme Schwarzschild-de Sitter black hole in \cite{cardoso}. In \cite{molina}, P$\ddot{o}$schl-Teller approximation was used to compute QNM for massive scalar field in d-dimensional Schwarzschild-de Sitter and Reissner-Nordstrom-de Sitter black hole at the near extreme limit.

Some hand waving arguments can be made by observing the expression for $\omega$ given above. When $m$ becomes large,  the dominant term for $V_0$ comes from $m^2$. Hence $\omega_R \approx \frac{\sqrt{\gamma}}{2}m + constant$ and $\omega_R$ will increase with  $m$. This was observed in Table 1.  On the other hand, it is clear that $\omega_I$ does not depend on $m$ and this was clear in Table 1.

When $l$ is large, one can substitute the expression for $V_0$ and expand for large $l$ to see how the QNM frequency $\omega$ behaves as follows:

\be \label{expansion}
\omega_{large \hspace{0.1 cm} l}  \approx l \sqrt{ \frac{ \gamma}{ 4 \rho^2}} - \frac{i}{2} \eta ( n + \frac{1}{2})
\ee
Hence as obvious from the above equation{$\refb{expansion}$, $\omega_R$ depends on $l$ linearly for large $l$ and this was observed in Fig.$\refb{omegasphe}$ when $\omega_R$  vs $l$ is plotted for large $l$. Again, for large $l$, $\omega_I$ does not depend on $l$ and was clear from Fig.$\refb{omegasphe}$.


\section{Conclusions and future directions}

Our main goal in this paper has been to study QNM   of   the dilaton-de Sitter charged black hole under scalar perturbations. The black hole has two physical horizons: cosmological horizon($r_c$),
and the black hole event horizon($r_h$).  The causal structure of the black hole space-time changes significantly due to the presence of the dilaton. There is a curvature singularity at a finite radius, $r = 2D$, compared to the Schwarzschild-de Sitter black hole.  

We have used the sixth order WKB approximation to compute QNM frequencies.  The parameters of the theory such as $D, \Lambda, l$ and $n$ were changed to see how QNM frequencies depend on them. When $D$ is increased, both $\omega_R$ and $\omega_I$ increased. $\omega_I$ seems to have a linear relation with $D$. Since $\omega_I$ is  large for large $D$, black hole with the dilaton is more stable. When $\Lambda$ increases, $\omega_R$ and $\omega_I$ decreases and they both appear to have a linear relation with $\Lambda$. Hence, smaller $\Lambda$ is preferred for stable black holes. When the spherical harmonic index $l$ is increased, $\omega_R$ increase linearly and $\omega_I$ decreases and approaches a constant value.  We also observed  that $\omega_I$ vary linearly against the temperature which is similar to what was observed for AdS black holes. We have done an analytical study of $l=0$ modes.

QNM of massive scalar perturbations were also studied.  We verified the non-existence of quasi resonance modes. When the mass of the field $m$ is increased, $\omega_R$ increased  and $\omega_I$ decreases. However, $\omega_I$ does not reach zero even if $m$ is increased to large values. This observation is in contrast to what happens in asymptotically flat black holes where for large $m$,  $\omega_I$ reaches zero leading to zero damping.

Finally, we demonstrated that at the near-extremal limit (when $r_h \approx r_c$), the scalar field equation will have the P$\ddot{o}$schl-Teller potential. Hence, one can get exact expressions for the QNM frequencies. Most of the observations we did with WKB analysis can be explained with these exact expressions.

A natural extension of this work is to study QNM of gravitational perturbations of the dilaton-de Sitter black hole. Ferrari et.al \cite{ferrari} studied the gravitational perturbations of the dilaton black hole with $\Lambda =0$ ( GMGHS black hole). They noticed that the dilaton black hole breaks the isospectrality of axial and polar perturbations which characterize both the Schwarzschild and Reissner-Nordstrom black hole. It would be interesting to understand how the presence of the cosmological constant change these observations.

It would be also interesting to study dS/CFT correspondence for the black hole we studied in this paper.

\vspace{0.5 cm}


{\bf Acknowledgments:}  The author wish to thank R. A. Konoplya for providing the Mathematica file for  WKB approximation.


\end{document}